\documentstyle[aps,twocolumn,psfig,colordvi]{revtex}
\def\Green#1{#1}\def\Red#1{}\def\Blue#1{}

\begin{document} 
\title{
Stabilizing distinguishable qubits against spontaneous decay
by detected-jump correcting quantum codes
}
\author{
G. Alber$^1$,
Th. Beth$^2$,
Ch. Charnes$^{2,3}$,
A. Delgado$^1$,
M. Grassl$^2$,
M. Mussinger$^1$
} 
\address{$^1$ Abteilung f\"ur Quantenphysik, Universit\"at
Ulm, D--89069 Ulm, Germany\\
$^2$ Institut f\"ur Algorithmen und Kognitive Systeme, Universit\"at Karlsruhe,
D-76128 Karlsruhe, Germany\\
$^3$ Department of Computer Science \& Software Engineering,
University of Melbourne, Parkville, Vic 3052, Australia
}
\maketitle

\begin{abstract} 
A new class of error-correcting
quantum codes is introduced capable of
stabilizing qubits against
spontaneous decay arising from couplings to
statistically independent reservoirs. These quantum codes
are based on the idea of using an embedded quantum code and
exploiting the
classical information available
about which qubit has been affected by the environment.
They are immediately relevant for quantum 
computation and 
information processing
using arrays of trapped ions or nuclear spins.
Interesting relations between these quantum codes
and basic notions of design theory are established.
\end{abstract}

Much of the newly emerging field of quantum information processing
is driven by the desire to push the
characteristic quantum effects
into the macroscopic domain as far as possible.  For this purpose it
is of vital interest to develop efficient methods
for stabilizing the coherence of quantum systems against
destructive environmental influences. 
Prominent examples of such environmentally-induced dissipative
phenomena are spontaneous decay processes
originating from couplings of a quantum system to uncontrollable
photonic \cite{master} or phononic \cite{Weiss} degrees of freedom.
So far,
various efficient quantum error-correcting strategies have been developed.
All of them rely on redundancy as far as the encoding of
information in quantum states is concerned.

In principle, any quantum system can be stabilized against 
environmental influences by active quantum error-correcting codes (QECCs)
with the help of repeated control measurements and appropriately
conditioned recovery operations
\cite{passive0,active1,active2,active3,active4,Knill}
and by exploiting the
quantum Zeno effect \cite{Zeno,Beige}.
But, typically these QECCs require a large number of measurement and recovery
operations.
In physical systems governed by collective spontaneous decay processes
originating from couplings to a single reservoir
it is more
advantageous to use quantum error avoiding codes (QEACs)
\cite{passive1,passive2,passive2a}
for efficient stabilization.
These QEACs rely on the existence of a sufficiently high dimensional
decoherence free 
subsystem (DFS) which stabilizes the dynamics passively without
measurements and recovery operations.
However, in the opposite dynamical regime of
interest in which the spacings between physical qubits are much
larger than the wave lengths of spontaneously emitted
photons or phonons, these qubits 
decay into statistically independent reservoirs \cite{master,passive0}.
Efficient error-correcting strategies have also been proposed
for these
situations. Typically
they use a QECC constructed within a DFS \cite{passive0,passive3,Plenio}.
Errors arising from the conditional time evolution between two
quantum jumps are corrected passively by the QEAC
while quantum jumps are corrected actively by the QECC.
In this way the total
number of control measurements and recovery operations required
is decreased significantly in comparison with purely active QECCs.
However, so far the redundancy of these
embedded
codes is not satisfactory.
Plenio et al. \cite{Plenio} have constructed an embedded code capable
of stabilizing one logical qubit which uses eight physical ones,
but had not found a shorter code.
Thus, in view of present day experimental possibilities \cite{Monroe
} 
it is desirable to develop alternative error-correcting
strategies for these cases by which it
is possible to reduce redundancy even further without
loosing the advantage of passive error correction between successive
quantum jumps.

Motivated by this need, 
in this letter a new class of  
embedded quantum error-correcting codes is introduced
which 
are capable of stabilizing 
distinguishable qubits against
spontaneous decay processes.
These codes are based on the idea of embedding
an active QECC
within
a passive QEAC
and simultaneously
exploiting the classical information available about which
qubit has been affected by an environment. 
Optimal
one detected-jump
correcting quantum codes 
of even length are constructed 
which minimize redundancy. It turns out that
their redundancy is significantly smaller than that of
previously proposed 
embedded error correction schemes \cite{Plenio}.
This latter property
makes these new quantum codes particularly attractive for
quantum computation and information processing
based on arrays of trapped ions or nuclear spins.
A 
link to basic notions
of design theory \cite{beth} is established which is expected to be 
useful for further explorations of this basic idea.

Let us consider $n$ distinguishable qubits which are perturbed by
statistically independent reservoirs 
inducing spontaneous decay processes.
Within the Markov approximation the time evolution of the density operator
$\rho$ of these
 $n$
qubits can be described by
a master equation
\begin{eqnarray}
&&\dot{\rho}(t)
=-\frac{i}{\hbar}[H,\rho(t)] + \frac{1}{2}\sum_{\alpha = 1}^{n}\{
[L_{\alpha},\rho(t) L_{\alpha}^{\dagger}] + 
[L_{\alpha}\rho(t), L_{\alpha}^{\dagger}] \}. 
\label{Lindblad}
\end{eqnarray}
Thereby the Lindblad operator $L_{\alpha} =
\sqrt{\kappa_{\alpha}}|0\rangle_{\alpha} \langle 1|_{\alpha}$ 
characterizes spontaneous decay of qubit $\alpha$ from
its excited state $|1\rangle_{\alpha}$
into its stable state $|0\rangle_{\alpha}$ with rate $\kappa_{\alpha}$ \cite{master}. 
The coherent part of the
$n$-qubit dynamics
is described by the Hamiltonian $H$.
In the case of radiative damping of quantum optical systems
the derivation of Eq.~(\ref{Lindblad}) 
involves the Born and the Markov approximations which are typically
very good. These approximations
rest on the assumption of 
weak couplings between resonantly excited two-level systems
and the vacuum modes of the electromagnetic field and a sufficiently
short correlation time of these vacuum modes \cite{master,passive0}.
However, in solid state devices where spontaneous decay
processes typically originate from couplings to phononic reservoirs
this Markov approximation is usually only applicable for sufficiently
high temperatures of the reservoirs \cite{Weiss}.
Within the quantum trajectory approach \cite{master}
the solution of Eq.~(\ref{Lindblad})
can be unravelled into a statistical ensemble of pure states. 
Each element of this ensemble defines a quantum trajectory
which describes the $n$-qubit system conditioned on 
the observation of $N$ quantum jumps
of qubits $\alpha_1,\ldots,\alpha_N$ which take place at times
$t_1\leq \ldots\leq t_N$. 
The action of these quantum jumps is represented by a sequence of
Lindblad operators $L_{\alpha_1},\ldots, L_{\alpha_N}$.
In this quantum 
trajectory
representation 
the conditional time evolution between two successive
quantum jumps is determined by the non-hermitian
effective Hamiltonian
$\tilde{H}=H - i(\hbar/2) \sum_{\alpha=1}^{n}L_{\alpha}^{\dagger}L_{\alpha}$.

The dynamics described by Eq.~(\ref{Lindblad}) can be stabilized against these dissipative 
influences in an effective way
by
embedded quantum codes.
For this purpose
one constructs first of all a DFS which stabilizes the conditional
time evolution 
between two successive quantum jumps passively.
In a second step one inverts the occurring quantum
jumps with the help of 
an active QECC which is
constructed within this DFS
\cite{passive0,passive1,passive2,passive3,Plenio}.
Thus, for this stabilization it is 
necessary to observe the $n$-qubit system continuously.
Whenever a quantum jump occurs the appropriate
unitary recovery operation is applied
within a time interval short
in comparison with the decay times and with the coherent evolution time
of the system \cite{passive0}.
The mean number of required
recovery operations is 
determined by the spontaneous decay
rates $\kappa_{\alpha}$ of the qubits.
For the simplest 
case possible of encoding
only one logical qubit 
Plenio et al. \cite{Plenio} have presented 
a one-error correcting
embedded quantum code 
which applies to the important special case of equal decay
rates of all the qubits.
Their active QECC constructed within the DFS
fulfills the conditions 
\begin{eqnarray}
&&
\langle c_i|L_{\alpha}^{\dagger}L_{\beta}|c_j\rangle =
\Lambda_{\alpha \beta} \delta_{ij}
\label{Knill}
\end{eqnarray}
for any qubits $\alpha$, $\beta$ and any logical states
$|c_i\rangle$, $|c_j\rangle$ with $\langle c_i|c_j\rangle = \delta_{ij}$.
These conditions are necessary and sufficient for the existence of
appropriate recovery operations  \cite{Knill} in all cases where 
an unknown qubit has been affected by a
quantum jump at a known jump time.
Being consistent with conditions (\ref{Knill})
Plenio et al.\cite{Plenio}
were not able to reduce the redundancy of their code any further.

In the subsequent treatment, however, it is demonstrated that the
redundancy of 
embedded quantum codes can be reduced significantly 
by also taking into account  the available information about which
qubit has been affected by a quantum jump. If the qubits of a quantum
computer couple to independent reservoirs 
then information about the jump time, say $t$, and about the jump
`position', say $\alpha$, is available.
Therefore, it is natural to exploit this additional information
about the `position' of a quantum jump for a more efficient  encoding.
If one can determine not only the jump time $t$
but also the jump position $\alpha$ by continuously monitoring
the $n$-qubit quantum system, one has to correct the error
operator $L_{\alpha}$ only for this particular value of $\alpha$.
As a consequence the corresponding active
QECC has to fulfill Eqs.~(\ref{Knill}) only for $\alpha=\beta$. 
The violation of conditions (\ref{Knill}) for $\alpha\neq \beta$
offers the possibility to construct
embedded codes
with a significantly smaller degree of redundancy. 
It should be mentioned that a similar violation of conditions (\ref{Knill})
has also been
realized previously in the treatment of the quantum erasure channel
\cite{erasure}.

Let us concentrate on
the important special case of equal spontaneous
decay rates of all the qubits, i.e.
$\kappa_{\alpha}=\kappa_{\beta} \equiv \kappa$.
If the number of physical qubits $n$ is even, the DFS
of maximal dimension with respect to the conditional time evolution
between successive quantum jumps is formed by all
$n$-particle
quantum states with $(n/2)$ excited and $(n/2)$ unexcited qubits.
This DFS is  
the eigenspace of the operator
$\sum_{\alpha=1}^{n}L_{\alpha}^{\dagger}L_{\alpha}$ with eigenvalue
$\kappa(n/2)$ and with dimension 
$d={n \choose n/2}\equiv n!/[(n/2)!]^2$.
Thus, 
the conditional
time evolution between successive quantum jumps is not perturbed by
the reservoirs.
Furthermore, for a given number of physical qubits $n$
the dimension of this DFS is maximal so that the degree
of redundancy is minimal.
For the correction of quantum jumps we have to develop an active QECC
within this DFS.
Thereby we want to exploit the 
fact that we have to correct quantum jumps only which take place
at a known `position', say $\alpha$.
Let us start with the simplest case possible,
namely the encoding of a single logical qubit.
We propose the following four-qubit encoding (omitting normalization)
\begin{eqnarray}
&&|c_0\rangle = |1100\rangle +|0011\rangle,\quad
|c_1\rangle = |0110\rangle +|1001\rangle
\label{four}
\end{eqnarray}
formed by complementary pairs within this DFS. Thereby
$|c_0\rangle$ and $|c_1\rangle$ encode the logical states $|0\rangle_L$
and $|1\rangle_L$. The complementary pairs appearing in Eq.~(\ref{four})
involve an excited state at any `position'.
Provided the error `position' $\alpha$ is known
this encoding represents an active QECC 
correcting $t=1$ detected-jump error and is formed by
superpositions of basis states of the DFS with
dimension $d=6$.
This encoding violates Eqs.~(\ref{Knill}) for $\alpha \neq \beta$ as
 $\langle c_0|L_4^{\dagger}L_2|c_1\rangle =
 \langle c_0|L_3^{\dagger}L_1|c_1\rangle \neq 0$, for example.
Provided a quantum jump $L_{\alpha}$ has occurred
at `position' $\alpha$ the immediate application of the
unitary recovery operator $R_{\alpha} =
\pi_{\alpha}(\prod_{\alpha \neq \beta}C_{\alpha \beta})\pi_{\alpha}H_{\alpha}$
restores the unperturbed
quantum state again. Here $\pi_{\alpha}$,
$H_{\alpha}$, and  $C_{\alpha \beta}$ represent a
$\pi$-rotation, 
a Hadamard transformation
of qubit $\alpha$, and a conditional $CNOT$ operation with
control and target qubits $\alpha$ and $\beta$. 
On the code space formed by all 
linear combinations of the logical
states of Eq.~(\ref{four}) $R_{\alpha}$ is the left-inverse of the quantum jump
operator $L_{\alpha}$. For the construction of
such a left-inverse unitary recovery
operator $R_{\alpha}$ \cite{Knill} the codewords have to fulfill the
necessary and sufficient conditions
\begin{eqnarray}
&&
\langle c_i |L_{\alpha}^{\dagger}L_{\alpha}|c_j \rangle
=\Lambda_{\alpha}\delta_{i j}.
\label{condition}
\end{eqnarray}
These conditions reflect the fact that the invertibility conditions
of Eq.~(\ref{Knill}) have to be fulfilled only for
$\alpha=\beta$. 
It should also be mentioned
that it is also possible to encode a third logical
quantum state $|2\rangle_L$ within the above mentioned DFS by the state
$|c_2\rangle = \frac{1}{\sqrt{2}}(|1010\rangle +|0101\rangle)$.
Thus, the three logical quantum states $|c_0\rangle,|c_1\rangle,|c_2\rangle$
represent a three-dimensional
one 
detected-jump correcting quantum code
formed by four physical qubits two of which are excited. Correspondingly 
we denote this code by $1$-JC$(4,2,3)$. 

It is straightforward to generalize  this construction to
arbitrary large numbers of logical states. In analogy to Eq.~(\ref{four}) one starts
from an even number $n$ of physical qubits and from the corresponding
DFS of dimension $d = {n \choose n/2}$. A basis of
this DFS consists of
all 
$n$-qubit states with $(n/2)$ excited and $(n/2)$
unexcited  
qubits. Within this DFS one forms the logical states of the
active QECC from all 
equally weighted complementary
pairs of states.
The resulting
embedded quantum code can correct 
$t=1$
detected-jump error. It
is optimal in the sense that
for a given number
$n$ of physical qubits the number of
logical 
states $l=\frac{1}{2}{n\choose n/2}$ is maximal.
Thus, for a large number of physical qubits $n$ the associated number of
logical qubits that can be encoded is given by
$\log_2 l = n - \frac{1}{2}\log_2 n + O(1)$.

The optimality of this encoding can be shown by
the following estimate of dimension.
For a given number $n$ of physical qubits with $k$ excited
states and a given number $t$ of errors
at known `positions' $\alpha_1,\ldots,\alpha_t$
the number of logical
states $l$ is bounded by the inequality
$l \leq {n - t \choose k-t}$. This upper bound originates from the fact that
after $t$ quantum jumps $t$
qubits are in state $|0\rangle$ at known `positions'. As the logical states have
to be recovered from these latter states by a 
unitary transformation the dimension of this latter Hilbert
space
also determines the maximum possible number of orthogonal logical states.
By the basic symmetry property of the binomial coefficients 
the maximum number of logical states is achieved for $k=[n/2]$.
($[x]$ denotes the largest integer smaller or equal to $x$.)
Thus we arrive at the final result that for $t=1$
the maximum number of logical quantum states is given by $l={n-1\choose n/2 -1}\equiv
\frac{1}{2}{n\choose n/2}$.

These one detected-jump correcting quantum codes
can be generalized
so that they are capable of correcting 
an arbitrary number $t$
of errors of an arbitrary number
of qubits.
Correspondingly, we define a
$t$ detected-jump
correcting quantum code, denoted by
$t$-JC$(n,k,l)$,
by a set of $l$ codewords $\{|c_i\rangle, i=1,\ldots,l \}$ formed by the linear
superpositions of $n$-qubit states each of which involves $k$ excited and $n-k$ unexcited states.
Analogous to Eqs.~(\ref{condition}) these codewords have
to fulfill the conditions
\begin{eqnarray}
\langle c_i|L^{\dagger}_{\bf e}L_{\bf e}|c_j\rangle =
\Lambda_{\bf e}\delta_{i j}
\label{JC}
\end{eqnarray}
which are necessary and sufficient for the existence of a unitary recovery operation.
Thereby the error operator $L_{\bf e}$ denotes an arbitrary product
of Lindblad operators,
say $L_{\alpha_m}\ldots L_{\alpha_1}$, corresponding to a jump pattern
${\bf e}\equiv (\alpha_1,\ldots,\alpha_m)$ of length $m$.
Eqs.~(\ref{JC}) have to be fulfilled for all 
jump patterns ${\bf e}$ of lengths $m$ not greater than $t$.
According to this terminology the previously
constructed optimal 
one 
detected-jump correcting quantum codes
are of the type $1$-JC$(n,n/2,\frac{1}{2}{n\choose n/2})$ with $n$ being even.
Furthermore, the above dimension estimate
implies that  
$t$-detected-jump correcting quantum codes
of the type
$t$-JC$(n,n/2,{n-t \choose n/2 -t})$ would be optimal.

The constructed one 
detected-jump correcting quantum codes
are particularly well suited for stabilizing quantum algorithms 
against spontaneous decay of the qubits into statistically independent reservoirs.
Thus, for example, they may be applied for stabilizing trapped-ion systems
\cite{Cirac} against radiative or for stabilizing
nuclear spin arrays \cite{Kane} against phononic damping
provided the mean distance between the
ions or spins representing the qubits is larger than the wave
lengths 
of the spontaneously
emitted photons or phonons. For this purpose one has to determine which qubit
has been affected by the spontaneous decay process. For spontaneously emitted photons, for
example, this may be achieved by photodetection techniques or by measuring the 
recoil of the affected particle.
This latter method may also be used in phononic decay processes.
Furthermore, one has to ensure
that in the absence
of errors the quantum system remains within the appropriate
DFS throughout the entire computation.
Recent investigations by Bacon et al. \cite{DFScalc} demonstrate that this latter
requirement
may be achieved with the help of suitably chosen universal quantum gates which
do not leave this DFS during their application. In solid state implementations
such gates may be realized by appropriately
tuning the coefficients of the Heisenberg-type exchange
terms \cite{DFScalc} by externally applied electric or magnetic fields.
Similarly, such a tuning appears also feasible for ions in arrays
of microtraps \cite{Cirac}
by applying appropriate laser pulses which push the ions out of their equilibrium
positions in a state dependent way.
At the time of writing this letter the
controlled manipulation of 
four qubits in ion traps 
seems to be in reach
\cite{Monroe}. Therefore, already the most
simple example of the presented
optimal one 
detected-jump correcting codes,
namely the $1$-JC$(4,2,3)$-code,
might give rise to interesting experimental settings.
We also want to point out that
strictly
speaking all the presented 
detected-jump correcting quantum codes
stabilize qubit
systems only with identical spontaneous decay rates for all
qubits. However, recent investigations on stability properties of
concatenated quantum codes indicate \cite{Lidar3} that these codes are
expected to stabilize also other qubit systems to a satisfactory
degree as long as all relative differences between spontaneous decay
rates remain small.

\begin{figure}
\centerline{\psfig{figure=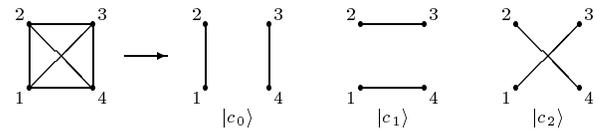,clip=}}
\medskip
\caption{Graphical representation of the affine plane of 4
points and 6 lines. The partition into 3 disjoint parallel classes of
lines defines the states of the $1$-JC$(4,2,3)$.}
\label{Affine}
\end{figure}

\begin{figure}
\centerline{\psfig{figure=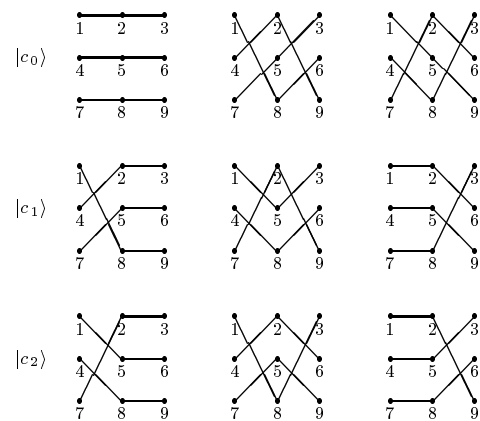,clip=}}
\medskip
\caption{This arrangement of 27 blocks of size 3 into disjoint rows of
3 parallel classes forms a $2$-SEED$(9,3,3)$. Superposition of the 9
blocks in each row \Green{of diagrams} yields the states
$|c_0\rangle$, $|c_1\rangle$, and $|c_2\rangle$ of a $2$-JC$(9,3,3)$.}
\label{SEED}
\end{figure}

For codewords consisting of linear superpositions of
quantum states 
with identical amplitudes
we can establish a surprising and far reaching connection with the area of
combinatorial design theory \cite{beth}.
This link 
seems to be particularly
fruitful for further explorations of general $t$-JC$(n,k,l)$-codes.
In order to exhibit basic ideas of this connection
let us finally reconsider the previously introduced optimal $1$-JC$(4,2,3)$-code.
Its three codewords $|c_0\rangle,|c_1\rangle,|c_2\rangle$
can be represented graphically
by the connected
diagram depicted in
Fig.~\ref{Affine}. Each
point of this diagram 
is
associated with a 
qubit. Two connected
points, i.e. a block, indicate that these two 
qubits are in the excited state $|1\rangle$.
Within the framework of finite geometry \Green{\cite{beth}} this connected
diagram
forms an affine finite plane over the binary field. In this context
the six blocks of
Fig.~\ref{Affine} represent lines,
i.e. one-dimensional subspaces of this geometry.
The three codewords $|c_0\rangle, |c_1\rangle,|c_2\rangle$
correspond to the three 
disjoint pairs of lines.
We call this combinatorial structure given by the partition of the set
of lines of Fig.~\ref{Affine} a 
$t=1$ spontaneous-emission-error
design,
$1$-SEED$(4,2,3)$, on $n=4$ points of
blocksize 
$k=2$ with $l=3$ disjoint
classes.
Generalizing this notion to arbitrary values of $(t,n,k,l)$ we arrive
at the notion of a $t$-SEED$(n,k,l)$.  
As an example, let us consider the $2$-SEED$(9,3,3)$ depicted in
Fig.~\ref{SEED}. Here the lines connecting 3 points indicate states
of 9 qubits in which 3 are excited. The sets of these points are
called blocks. Superposition of the 9 blocks of size 3 contained in
the 3 parallel classes depicted by any of the 3 rows of diagrams in
Fig.~\ref{SEED} gives the three codewords $|c_0\rangle$,
$|c_1\rangle$, $|c_2\rangle$ of a $2$-JC$(9,3,3)$, e.g.,
$|c_0\rangle = |111000000\rangle + |000111000\rangle + |000000111\rangle +
              |100001010\rangle + |010100001\rangle + |001010100\rangle +
              |100010001\rangle + |010001100\rangle + |001100010\rangle$.
For the construction of an arbitrary $t$-SEED$(n,k,l)$ design theory
\cite{beth} offers powerful combinatorial methods which will be
described in a subsequent article.

In summary, a new class or error-correcting quantum codes has been introduced
for stabilizing qubits against spontaneous decay into independent reservoirs.
It is based on the idea of using embedded quantum codes and simultaneously
exploiting classical information about the error position. 
Thus, redundancy can be reduced significantly. 
The systematic 
construction and classification of
$t$-JC$(n,k,l)$-codes with $t\geq 2$ which minimize redundancy is
still a challenging task which is currently under
active investigation.  Here the
newly discovered
relation to design theory seems to play a
key role, especially for the construction of
$t$-SEEDs with large $t$.

This work is supported by the DFG (SPP~1078) and by the EC
(IST-1999-10596). The work of A.D. is also supported by the DAAD.

\end{document}